\documentclass[reprint]{JASA}

\usepackage{musixtex}
\usepackage{multirow}
\usepackage{booktabs}
\usepackage{siunitx}
\usepackage{array}
\usepackage{CJKutf8}

\begin{document}
\title[JASA/Sample JASA Article]{Acoustic Characterization of the Resonator in the Chinese Transverse Flute (dizi)}
\author{Xinmeng Luan}
\email{xinmeng.luan@mail.mcgill.ca}
\author{Song Wang}
\author{Gary Scavone}
\affiliation{Computational Acoustic Modeling Laboratory, Center for Interdisciplinary Research in Music Media and Technology, Schulich School of Music, McGill University, Canada}

\author{Zijin Li}			
\affiliation{Department of Music AI, Central Conservatory of Music, China}

\preprint{Author, JASA}	

\date{\today}

\begin{abstract}
The \textit{dizi} is a traditional Chinese transverse flute and is most distinguished from the western flute by the presence of a hole covered by a wrinkled membrane. In this study, we analyze the linear acoustical behavior of the \textit{dizi} resonator through a detailed acoustical model that incorporates drilled toneholes, back end-holes, membrane hole, and upstream embouchure hole. The input admittance of the \textit{dizi} is measured and modeled using the Transfer Matrix Method (TMM) and Transfer Matrix Method with external Interactions (TMMI). In comparison to measurements, the TMMI is shown to more accurately model the \textit{dizi} than the TMM when compared to measurements. Our analysis reveals that attaching the membrane shifts admittance peaks to lower frequencies, reduces their magnitude, and influences tuning and harmonicity for different peaks and fingerings. The study further shows that the upstream branch, which includes the embouchure hole, complicates the evaluation of the tonehole lattice cutoff frequency, suggesting that it may not need to be considered for flute instruments. Cutoff frequencies exhibit distinct groupings across fingerings, influenced by the different tonehole lattices in the \textit{dizi}: finger-hole lattice and end-hole lattice.


\end{abstract}


\maketitle


\section{\label{sec:1} Introduction}
\begin{figure}
    \centering
 \includegraphics[width=1\linewidth]{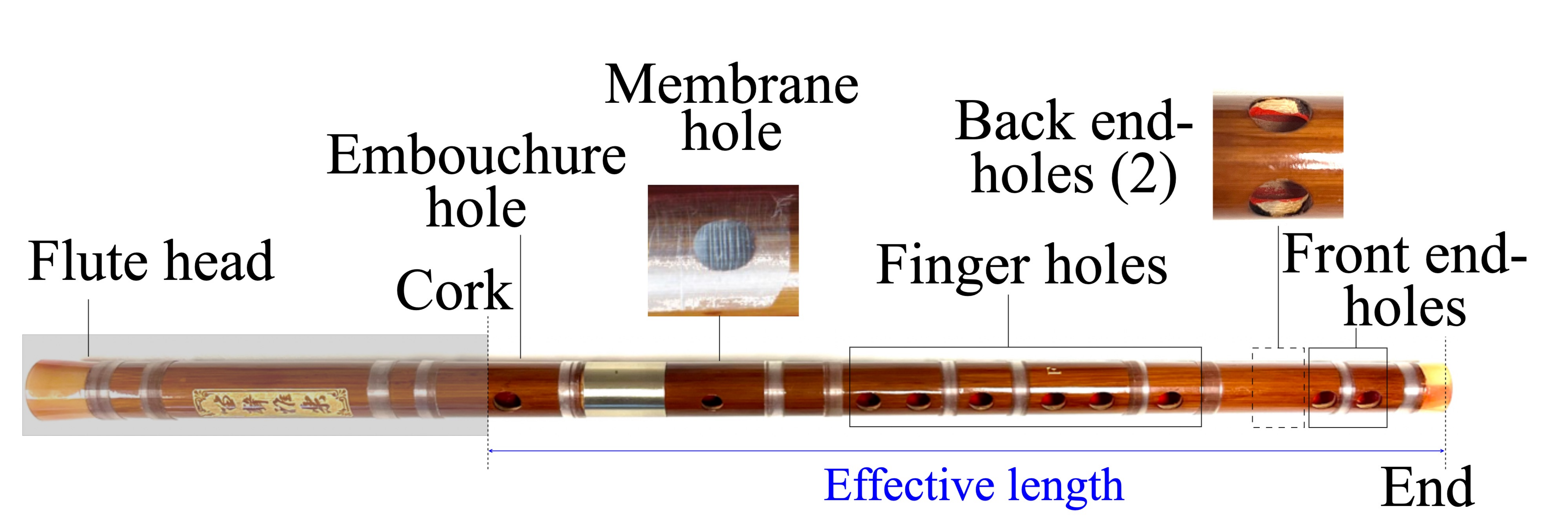}
        \caption{Image of the F-key \textit{bangdi} and labeled parts.}
    \label{fig:generaldizi}
\end{figure}

The \textit{dizi} (\begin{CJK}{UTF8}{gbsn}笛子\end{CJK}), a traditional Chinese transverse flute crafted primarily from bamboo, is most distinguished from the western flute by the presence of a hole covered by a wrinkled membrane. The wrinkled membrane is known to contribute to the unique sound brightness of the \textit{dizi}. 
The general shape of the \textit{dizi}, shown in Fig.~\ref{fig:generaldizi}, is similar to that of the western flute, featuring a primarily cylindrical bore and a series of toneholes along its length. The membrane hole is located between the embouchure hole and six downstream finger-holes. 
Four extra, non-fingered end-holes are located near the bottom of the bore as follows: two axially distributed front end-holes, like the other toneholes along the front side, as well as two radially distributed back end-holes. 
Similar configurations of end-holes can also be found on the \textit{Xiao}, a Chinese longitudinal flute. 
An internal cork, not visible externally, is located a short distance above the embouchure hole. Thus, the acoustically-relevant section of the instrument is from the cork to the downstream end. 
The upstream portion from the cork to the end identified as the ``flute head" in Fig.~\ref{fig:generaldizi} is not modeled in this study. 
It is worth noting that the membrane is a unique component of the \textit{dizi}, crafted from a thin film sourced from reed or bamboo stems, which makes it particularly delicate and fragile. The membrane's lifespan is short, typically lasting only a few months at most. There are various aspects that need to be taken into account when selecting a suitable membrane, such as the type of the \textit{dizi}, playing techniques, and music style, as it significantly affects the instrument's timbre. 
Before playing the \textit{dizi}, the musician needs to manually create wrinkles on the membrane and attach it to the membrane hole, as shown in Fig.~\ref{fig:generaldizi}.
Furthermore, the method of attaching the membrane, its natural texture, and the formation of wrinkles, make a difference of tonal quality.

The fingering technique of the \textit{dizi} includes half-holing, which involves partially covering a finger-hole, typically about 1/3 to 1/4 of its area. Proper tuning with this technique requires precise finger positioning and ear training. 
Additionally, cross-fingerings are used for certain notes, particularly in fast musical passages. 
Players can extend the range of the instrument into the second and third octave by overblowing. 
The commonly used keys for the \textit{dizi} are C, D, E, F, and G. \textit{Dizi} in the keys\footnote{The naming convention for the key of a \textit{dizi} is such that the note produced with all finger-holes covered corresponds to the dominant or major fifth of the key. Thus, for a \textit{dizi} in the key of F, a C5 (523.25 \si{\hertz}) sounds when all finger-holes are covered.} of C, D, and E are longer and thicker, making them more demanding in terms of breath control, and are referred to as \textit{qudi} (\begin{CJK}{UTF8}{gbsn}曲笛\end{CJK}). In contrast, \textit{dizi}  in F and G keys, known as \textit{bangdi} (\begin{CJK}{UTF8}{gbsn}梆笛\end{CJK}), are shorter, thinner, and require less breath to play. 

The physical principles that govern the production and radiation of sound by western musical instruments have been explained in detail in \citeauthor{Benadebook} (\citeyear{Benadebook}), \citeauthor{rossing1991physics} (\citeyear{rossing1991physics}) and \citeauthor{Chaigne} (\citeyear{Chaigne}).
However, there have been only a few scientific studies of Chinese musical instruments, with most of those coming in the last few decades. Some noteworthy contributions related to the \textit{dizi} are as follows:
\citeauthor{Tsai} (\citeyear{Tsai}) made important contributions to the study of the \textit{dizi} from both a physics and perception perspective. He explained the importance of the tension and wrinkles in the membrane in producing a beautiful, bright \textit{dizi} timbre.
The non-linear behavior of the membrane was studied by modeling it as a Duffing oscillator. 
\citeauthor{Tsai} (\citeyear{Tsai}) focused on a C-key \textit{qudi} for this study.
In another study, \citeauthor{ng} (\citeyear{ng}) measured and analyzed the input impedance of a G-key
\textit{bangdi}. They investigated the influence of a membrane on the \textit{dizi}'s resonances, finding that it has minimal effect, with only slight deviations for certain notes. They also compared different alternate fingerings for the same note, such as cross-fingering and half-holing, and discussed how the observation of impedance descriptors can impact playability from the musician's perspective.


A common approach to study the resonator of wind instruments, particularly the flute, involves one-dimensional modeling of the air column to obtain the input admittance, a linear acoustic frequency-domain characterization relating pressure and volume velocity at the embouchure hole.
The input admittance can be computed by the Transfer Matrix Method (TMM) \cite{Causse, Keefe1990} and the Finite Element Method (FEM) \cite{ernoult}.
The Transfer Matrix Method with external Interactions (TMMI) was proposed to incorporate the mutual radiation effect among openings based on the TMM \cite{tmmi}, which has been used to model saxophones \cite{tmmi}, and clarinets \cite{tmmi, petersen2020link}.

Employing input admittance as a tool for woodwind instrument analysis holds potential, as it provides information about both the magnitudes and harmonicity of resonances. 
Moreover, leveraging input admittance analysis aids in evaluating the influence of subtle alterations in hole and bore geometries across different fingerings, which can be used for instrument geometry optimization \cite{ernoult2020woodwind}. 
The input admittance can also be used for wind instrument sound synthesis \cite{taillard2018modal}.

Toneholes are the distinguishing feature of woodwind instruments.
The air column inside the main bore, along with the air in the toneholes, is referred to as the tonehole lattice.
In most contexts, the tonehole lattice refers to the open tonehole group excluding the first upstream open hole, which predominantly determines the playing frequency. The remaining toneholes, namely the open tonehole lattice (oTHL) \cite{petersen2021}, serve a secondary function by providing an effective length correction.
The cutoff frequency of an oTHL is a feature associated with wave propagation in periodic media, which produces pass and stop bands for a wave entering the lattice \cite{benade1988clarinet, benade1988saxophone}. 
This cutoff frequency is related to the tone color of the instrument, which can be reflected in the spectral characteristics of the radiated sound \cite{benade1988clarinet, benade1988saxophone}.

Selecting a high-quality \textit{dizi} is a challenge for professional musicians. Due to the organic nature of bamboo, craftsmen are constrained by the inherent characteristics of each bamboo blank, such as its internal and external dimensions, taper, uniformity, surface conditions, stiffness, and density, all of which vary from one piece to another \cite{ng}. This variability makes it difficult for craftsmen to achieve precise intonation when crafting a \textit{dizi}. The process of shaping the bamboo blank within design parameters to ensure consistent performance and reliable interaction with the player relies heavily on experience and ear training. There is a lack of scientific guidelines for the instrument making.

The first objective of this paper is to present an acoustic Transfer Matrix (TM) model for calculating the input admittance of the \textit{dizi} based on its geometry parameters. This model offers instrument makers a quick and scientific means to gain insights into the acoustics of the \textit{dizi}, facilitating the design process. The proposed model is validated against measurement results, demonstrating strong agreement with the TMMI model. 
Furthermore, the second goal is to perform a linear acoustic analysis of the \textit{dizi} through both measurements and modeling, considering both the instrument as a whole and its individual components. This analysis provides valuable insights into the tuning and playability characteristics of the instrument. 
Note that half-holing, which involves partially covering a tonehole based on the musician’s auditory judgment, is excluded from this study due to its inherent variability.

The paper is structured as follows.
First, the measurement system and process are discussed in Sec. \ref{sec:Measurement}. 
Then, the modeling details are extensively elaborated in Sec. \ref{sec:Modeling}. 
Section \ref{sec:Discussion} presents the discussion of the analysis of the measurement and modeling results. 
Finally, Sec. \ref{sec:conclusion} draws some final conclusions.

\section{\label{sec:Measurement} Measurement}
	
\begin{figure}
    \centering
        \includegraphics[width=0.9\linewidth]{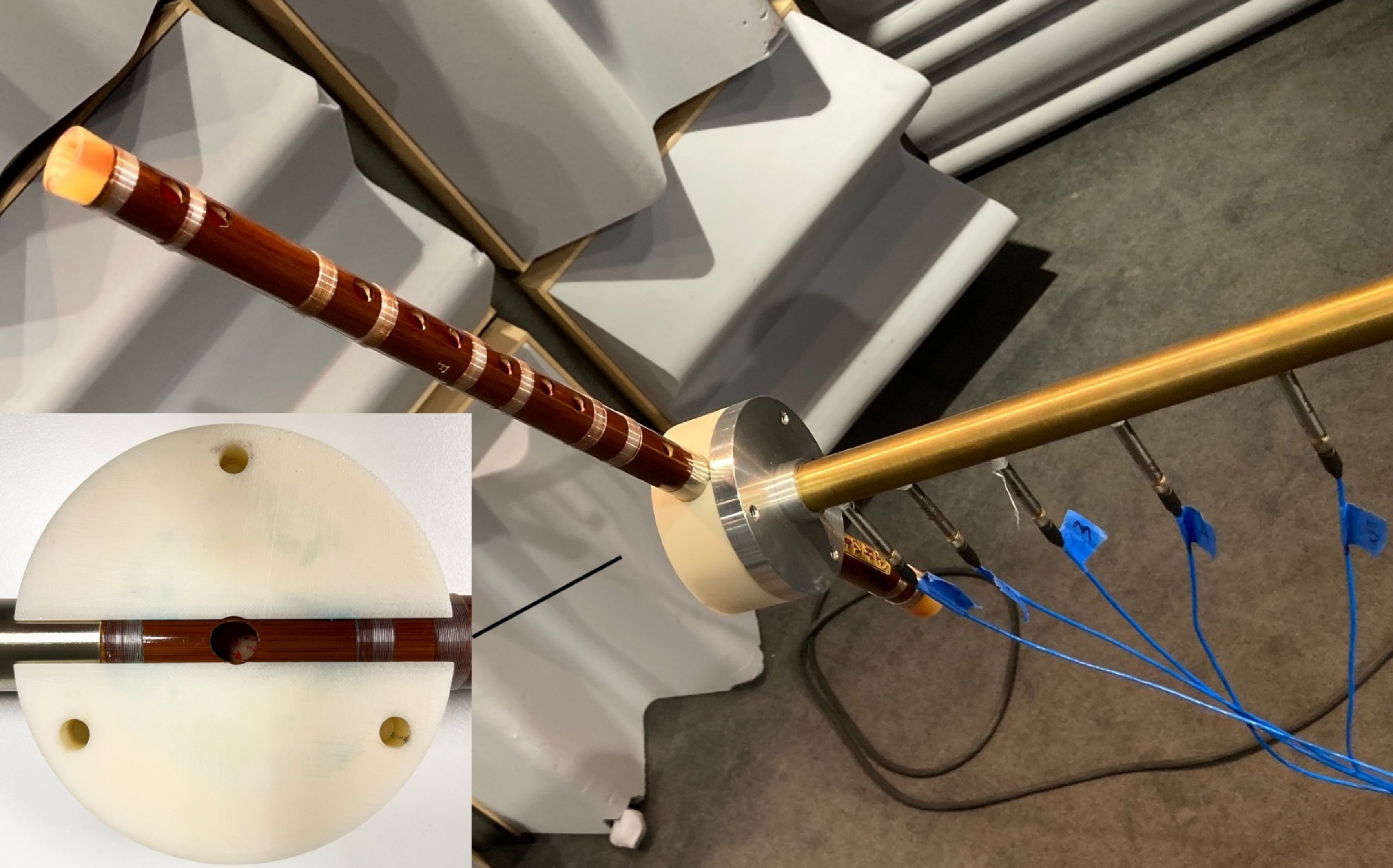}
        \caption{Image of the \textit{dizi} attached to the impedance measurement probe.}
      \label{fig:measys}
\end{figure}

A custom-build multi-microphone system based on a least-squares signal processing technique \cite{Lefebvremeasure,song} is used to measure the input admittance of the \textit{dizi} (see Fig.~\ref{fig:measys}). Six microphones are spaced along the impedance tube and three non-resonant loads are used to calibrate the apparatus, including a quasi-infinite impedance, an almost purely resistive impedance, and an unflanged pipe radiation load, similar to that described in \citeauthor{dickens2007} (\citeyear{dickens2007}) and \citeauthor{kemp2010} (\citeyear{kemp2010}).

The input admittance of each fingering is measured sequentially with and without the membrane. 
Blu-Tack is used to seal the toneholes and membrane hole (if the membrane is absent). When the membrane is present, it is attached in a wrinkled state, as it would be during a performance.
Seven different fingerings were measured, specified as XXXXXX, XXXXXO, XXXXOO, XXXOOO, XXOOOO, XOOOOO, OOOOOO, where O stands for open and X for closed, and the corresponding finger-hole sequence starts from the upstream end of the \textit{dizi}, corresponding to the notes C5 (\SI{523.25}{\hertz})/C6, D5/D6, E5/E6, F5/F6, G5/G6, A6/A7, B6/B7, respectively. 


The embouchure hole is connected to the reference
plane of the measurement system by a 3D-printed coupler (see the bottom left corner of Fig.~\ref{fig:measys}).
The dimensions of the \textit{dizi} are listed in Appendix \ref{app:dim}.
Note that only the linear behavior of the \textit{dizi} is captured due to the assumptions of the measurement approach and the use of a small-amplitude sound source excitation. 
For the case of large jet flow in actual performance, nonlinear characteristics will likely be present though such behaviour is not considered in this study.


\section{\label{sec:Modeling} Modeling }

\subsection{\label{subsec:tmmtmmi} Theory of TMM and TMMI}

The air column of a wind instrument can be modeled by
approximating its geometry as a series of segments. These segments can take the form of cylinders, cones, and toneholes (either closed or open). Each segment is characterized by a TM that establishes the relationship between the input and output frequency-domain parameters, specifically pressure and volume velocity. The overall response of the entire structure is then determined by multiplying the individual TMs in cascade.
The impact of the bore’s surface condition on the impedance spectra was investigated by \cite{boutin2017}; however, this aspect is not considered in our study.
  
 Using $\mathbf{T}_i$ to represent the TM of the $i$th element, the system containing $M$ elements can be expressed as
\begin{equation}	
\begin{bmatrix}
	p_{in}(\omega)\\u_{in}(\omega)
\end{bmatrix}=\left(\prod\limits_{i=1}^{M}{\mathbf{T}_i}(\omega)\right)\begin{bmatrix}
	p_{out}(\omega)\\u_{out}(\omega)	
	\end{bmatrix},	
\end{equation}
with $p_{in}(\omega)$, $u_{in}(\omega)$ and $p_{out}(\omega)$, $u_{out}(\omega)$ denoting input and output pressure and volume velocity as functions of the radian frequency $\omega$.
For simplicity, the frequency variable will be omitted in the following notation.
$Y_{in} = u_{in}/p_{in}$ represents the input admittance. 

For a cylindrical duct, the TM is
\begin{equation}
\mathbf{T}_{cyl}=\begin{bmatrix}
	\cosh(\Gamma L)&Z_{c}\sinh(\Gamma L)\\Z_{c}^{-1}\sinh(\Gamma L)&\cosh(\Gamma L)
\end{bmatrix},
\end{equation}
where $L$ is the length of the duct, $Z_c = \rho c/s$ is the characteristic impedance \footnote{Strictly speaking, the expression for $Z_c$ is an approximation, but it is sufficiently accurate for our modeling purposes.}, $c$ is the speed of sound, $s$ is the cross-sectional area, and $\Gamma$ is a complex propagation constant. $Z_c$ and $\Gamma$ depend on the acoustical constants of the gas and the duct diameter \footnote{The expression of $\Gamma$ is according to  \citeauthor{van2005} (\citeyear{van2005}). The thermodynamic constants for wave propagation in air are given in \citeauthor{keefe1984} (\citeyear{keefe1984}). The following default physical parameters are used in the modeling: the room temperature is \SI{20}{\degreeCelsius}, the relative room humidity is $40\%$ and the atmospheric pressure is  \SI{1013}{\hecto\pascal}.}.  

The elements of the TM of a short conical duct $\mathbf{T}_{cone}$ are \cite{Chaigne, tmmi}
\begin{equation}	
\begin{split}
\mathbf{T}_{cone}^{11}=&\frac{r_2}{r_1}\cos(k_cL)-\frac{\sin(k_cL)}{k_cx_1}, 
\\
\mathbf{T}_{cone}^{12}=&jZ_c\sin(k_cL), \\
\mathbf{T}_{cone}^{21}=&\frac{1}{Z_c} \bigg[ j\sin(k_cL) \left( 1 + \frac{1}{k_c^2x_1x_2} \right)  +  
 \frac{ \cos(k_cL)}{jk_c} \left( \frac{1}{x_1} - \frac{1}{x_2} \right) \bigg], 
\\
\mathbf{T}_{cone}^{22}=&\frac{r_1}{r_2}\cos(k_cL)  + \frac{\sin(k_cL)}{k_cx_2}.
\end{split}
\end{equation}
where $r_{1}$ and $r_{2}$ are the radii at the input and output planes, respectively, and $x_{1}$ and $x_{2}$ are the distances between the apex of the cone and the input and output planes, $Z_{c} = \rho c/(\pi r_{1}r_{2})$ and $k_{c}=-j\Gamma $ is the complex wavenumber.
In this case, losses are evaluated at the equivalent radius \cite{Chaigne}
\begin{equation}
    r_{eq}=L\frac{r_{1}}{x_{1}}\frac{1}{\ln {(1+L/x_{1})}}.
\end{equation}

\begin{figure}
    \centering
    \includegraphics[width=0.8\linewidth]{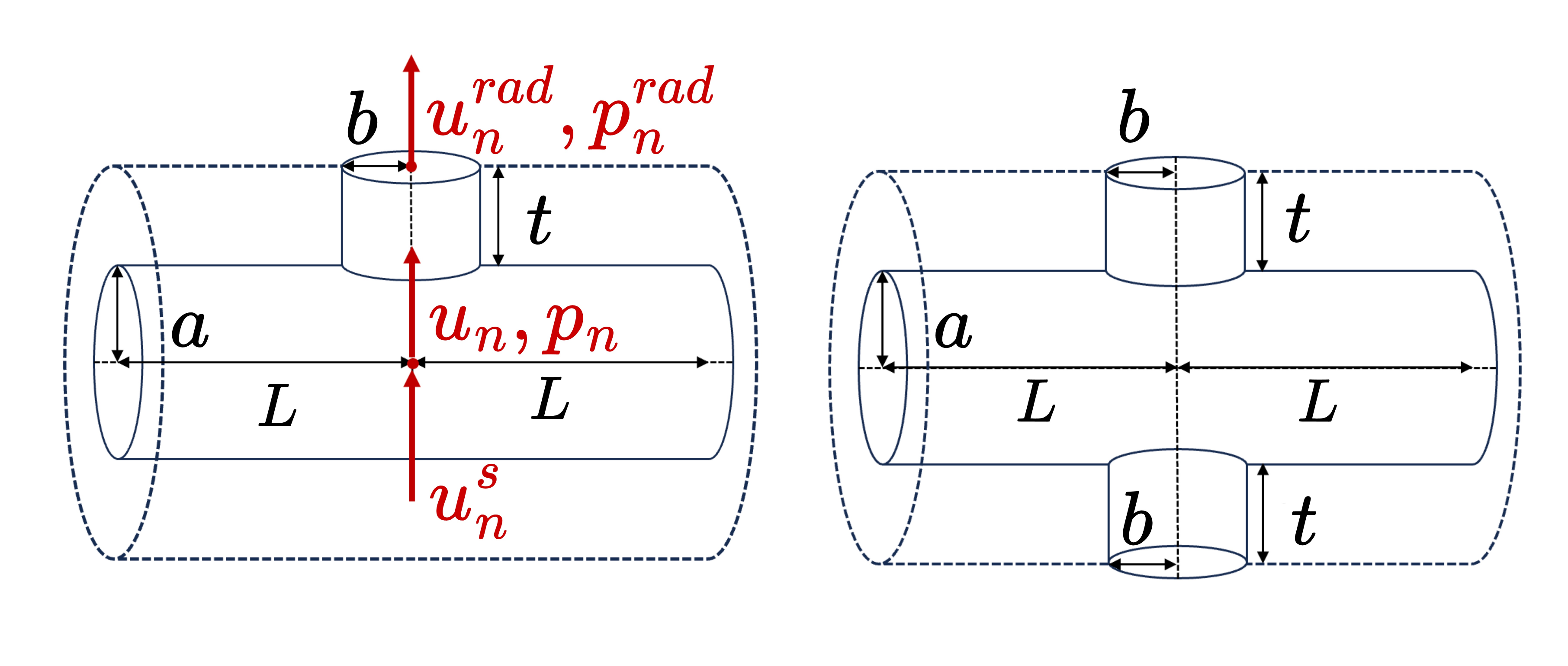}
    \caption{Left: Single tonehole. Right: Back-end hole.}
    \label{fig:SingleHole}
\end{figure}

The TM of a woodwind instrument's tonehole (Fig.~\ref{fig:SingleHole}) is written as \cite{keefe1982}
\begin{equation}
\begin{split}
	\mathbf{T}_{hole} &=\begin{bmatrix}
	1 & Z_a/2\\ 0 & 1
\end{bmatrix} \begin{bmatrix}
	1 & 0\\ 1/Z_s & 1
\end{bmatrix}\begin{bmatrix}
	1 & Z_a/2\\0 & 1	
\end{bmatrix} \\
&=\begin{bmatrix}
	1+\frac{Z_a}{2Z_s}& Z_a(1+\frac{Z_a}{4Z_s})\\ 1/Z_s & 1+\frac{Z_a}{2Z_s}
\end{bmatrix}.	
\end{split}
\label{eq:Thole}
\end{equation}
The series and shunt impedances, $Z_a$ and $Z_s$, can be expressed in terms of equivalent lengths, which can be found in Appendix \ref{zazs}.
\citeauthor{dubos} (\citeyear{dubos}) provided an alternative expression for the shunt term, suggesting that $Z_s$ can be expressed as $Z_s - Z_a/4$. However, in this study, we utilize the model proposed by \citeauthor{keefe1982} (\citeyear{keefe1982}) in Eq.~\eqref{eq:Thole}.

For woodwind instruments, there are often open holes in close proximity to one another. The TMM does not account for possible external sound interactions between holes. The TMMI was proposed to incorporate the mutual radiation effect among openings, including open holes and the open end \cite{tmmi}.
The routine for the TMMI involves dividing the bore into two portions at the most upstream open hole. The impedance of the downstream portion with open holes is calculated as described below and then used as the load impedance for the upstream section, the impedance of which is calculated using the TMM.


The pressure and volume velocity within the $N$ openings (open holes and end) of the downstream portion where TMMI is applied, under the assumption that the height of the tonehole is smaller than the wavelength, can be expressed as 
\begin{equation}
    \textbf{p} = \textbf{p}^{rad}+ \textbf{B} \textbf{u} = \textbf{Z}\textbf{u}^{rad}+\textbf{B}\textbf{u}=(\textbf{Z}+\textbf{B})\textbf{u},
    \label{eq: tmmi}
\end{equation}
where the vectors are defined as $\textbf{p} = [p_1, ..., p_N]$, $\textbf{u} = [u_1, ..., u_N]$, $\textbf{p}^{rad} = [p^{rad}_1, ..., p^{rad}_N]$ and $\textbf{u}^{rad} = [u^{rad}_1, ..., u^{rad}_N]$.
The matrices $\textbf{B}$ and $\textbf{Z}$ have elements $B_{ij}$ and $Z_{ij}$, where $i \in [1,N]$ and $j \in [1,N]$. 
$\textbf{p}$, $\textbf{u}$, $\textbf{p}^{rad}$, and $\textbf{u}^{rad}$ represent the internal and external pressures and volume velocity rates at the openings, respectively.
For a tonehole, they refer to the pressure and volume velocity inside the air column and chimney radiation surface, respectively, as shown on the left in Fig.~\ref{fig:SingleHole}.
$\textbf{B}$ is a diagonal matrix that represents the impedance of the acoustic mass of the open holes, as seen in Appendix \ref{app:tmmi}.
$\textbf{Z}$ is the radiation impedance matrix, where the diagonal elements are self-impedances and the off-diagonal elements are mutual impedances.

 The volume velocity can be calculated by
\begin{equation}
    \textbf{u}=[\textbf{I}+\textbf{Y}(\textbf{Z}+\textbf{B})]^{-1}\textbf{u}^s,
\end{equation}
where $\textbf{I}$ is the identity matrix and $\textbf{Y}$ is the admittance matrix, as seen in Appendix \ref{app:tmmi}. Both $\textbf{I}$  and  $\textbf{Y}$ have the same size as  $\textbf{B}$ and  $\textbf{Z}$.
A flow-source vector $\textbf{u}^s = [u^s_1, ..., u^s_N]$  is introduced at each open hole for the calculation, which can be regarded as a virtual source. When we are only concerned with the input admittance calculation, the virtual source is just a choice of reference. One solution is to apply only a reference volume velocity to the left of the uppermost open hole. A pressure source is also a viable option.


The difference between the results of the TMM and TMMI becomes apparent as the number and size of open holes increases and the frequency becomes higher. Therefore, using TMMI can provide more accurate results, especially when considering frequency ranges above the cutoff frequency of the tonehole lattice.  
The modeling of the \textit{dizi} using TMM and TMMI in this work is performed using the Matlab toolbox \textit{acmt} \cite{scavone2024open}\footnote{\href{https://github.com/garyscavone/acmt}{https://github.com/garyscavone/acmt}}.

\subsection{\label{subsec:drilledhole}Matching volume length correction for the drilled toneholes}

The matching volume length correction in Eq. (\ref{tm}) is derived from the holes that extend beyond the wall of the bore, as in flutes and saxophones. However, the toneholes of the \textit{dizi} are directly drilled into a thick wall, referred to as ``drilled holes'' below, similar to recorders. Thus, the matching volume is slightly different \cite{lefebvrethesis, rucz2015}. 
\citeauthor{lefebvrethesis} 
 (\citeyear{lefebvrethesis}) calculated the total equivalent length of the closed drilled hole, including the modified matching volume length correction, which is merged with the other length corrections and cannot be separated.

\begin{figure}
    \centering
    \begin{minipage}{0.6\linewidth}
        \includegraphics[width=1\linewidth]{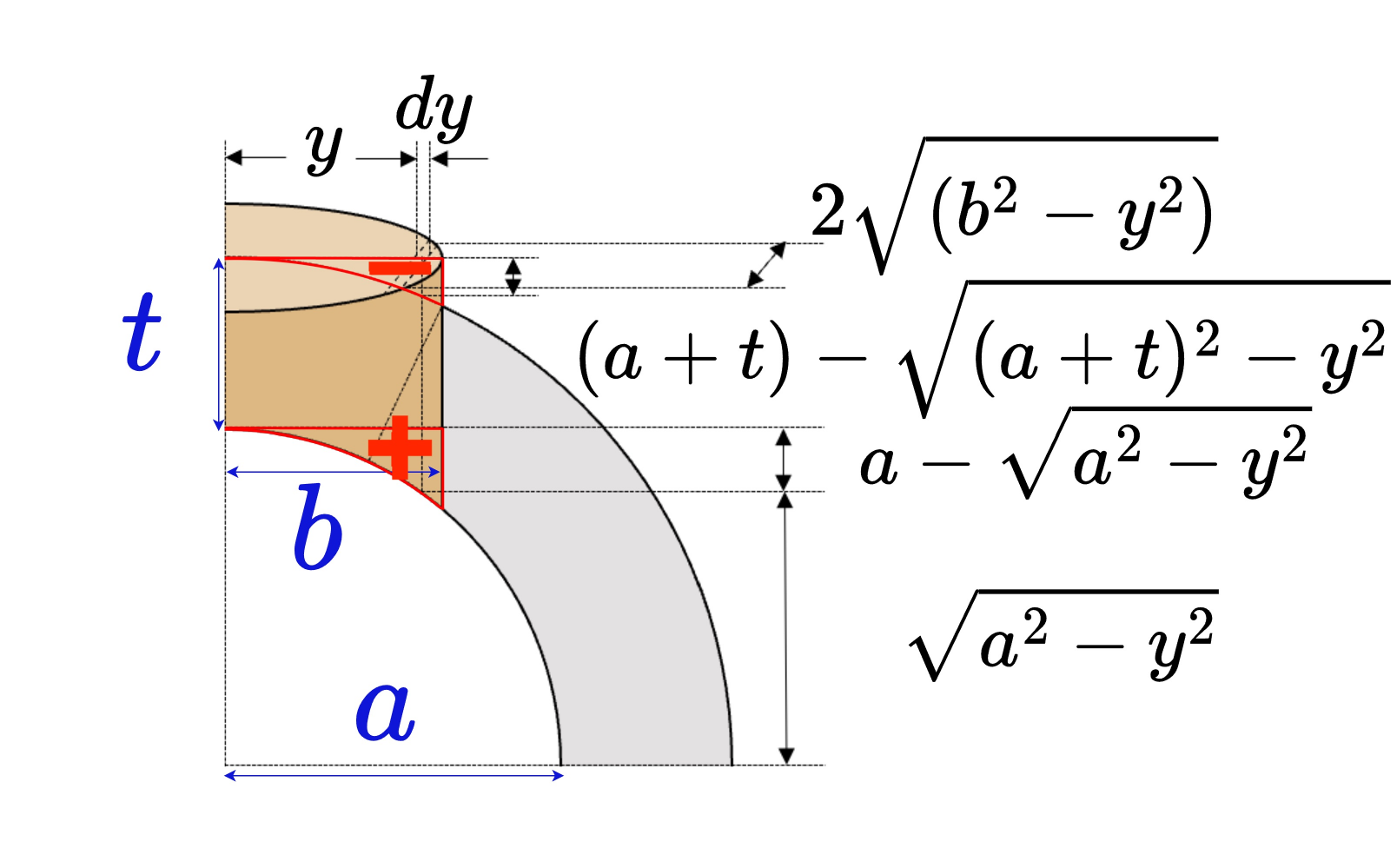}
    \end{minipage}%
    \hspace{1ex}
    \begin{minipage}{0.3\linewidth}
        \caption{\label{matching}Profile of a tonehole drilled in the thick wall.}
    \end{minipage}
\end{figure}

Figure \ref{matching} is a section view of the drilled hole. 
There are two volumes that need to be properly accounted for in defining the drilled tonehole parameters, including the addition of a region marked with a red $+$ and the subtraction of a region indicated by a red $-$ in the figure. In contrast, only the $+$ region is considered for extended holes \cite{nederveen}.
Note that a further volume correction term could be taken into account during performance, as the finger penetrates some distance into the hole.
According to the geometric illustration of the drilled hole in Fig.~\ref{matching}, the matching volume can be calculated by integrating as 
\begin{equation}
	V_m^+=\int _{-b}^{+b}2\sqrt{b^2-y^2}\left(a-\sqrt{a^2-y^2}\right)dy,
\end{equation}
and
\begin{equation}
	V_m^-=\int _{-b}^{+b}2\sqrt{b^2-y^2}\left(a+t-\sqrt{(a+t)^2-y^2}\right)dy,
\end{equation}
where $a$ is the bore radius, $b$ is the tonehole radius and $t$ is the tonehole height or wall thickness.
Then the total matching volume length correction can be expressed as
\begin{equation}	
t_m^{(o/c)}=\frac{1}{\pi b^2}(V_m^+- V_m^-).
\end{equation}
Through numerical integration, it can be transformed into 
\begin{equation}
	t_m^{(o/c)} = \frac{tb^2}{8a(a+t)} + o(b^5),
 \label{eq:match}
\end{equation}
where $o(b^5)$ indicates that the approximation is accurate to an order smaller than $b^5$.

\subsection{\label{subsec:3:3}Back end-holes}
The back end-holes refer to two holes distributed radially along the \textit{dizi}, as shown on the right in Fig.~\ref{fig:SingleHole}. 
 These holes present a challenge when developing a one-dimensional model, as they are co-located along the dimension of interest.
A similar structure could also be found on the \textit{Xiao}, another Chinese air-jet driven woodwind instrument. When using TMM to model the \textit{Xiao} in \citeauthor{Lan2016} (\citeyear{Lan2016}), the end-holes and the open end were treated together as a radiation impedance, which was obtained through measurement instead of modeling. The drawback of this approach is that it requires a measurement for each new instrument geometry.

Instead, the back end-holes can be approximated using an equivalent acoustic lumped model with a TM. As the dimensions of the two holes are nearly identical, they can be modeled as two parallel toneholes with the same parameters, which means doubling the series impedance and halving the shunt impedance. 
However, from a comparison of the measurement and modeling results, an empirical factor of 1/2.2 for the shunt impedance was found to produce a better fit. The use of this factor is substantiated given the lack of any precise analysis of this hole configuration and also because mutual radiation between the two holes, which is taken into account, will likely be lower for this hole configuration compared to two axially adjacent holes.



\subsection{\label{subsec:3:4} Membrane hole}
The equations of motion and the acoustic impedance for the \textit{dizi} membrane are derived by \citeauthor{Tsai} (\citeyear{Tsai}).
The wrinkled membrane sealing on the tonehole can be modeled as a mass-spring system and its impedance can be expressed as
\begin{equation}
	Z_{mem}=\frac{1}{S_m^2}\left [R_m+\frac{jm(\omega^2-\omega _m^2)}{\omega}\right ],
\end{equation} 
where $m$ is the membrane mass, $R_m$ is its mechanical damping coefficient and $\omega _m = 2 \pi f_m$ is the resonant radian frequency. Then the membrane hole can be modeled using a closed-hole TM.  In this model, the closed-hole shunt impedance is considered in series with the membrane impedance, which captures the acoustic behavior of the membrane without accounting for mutual radiation interactions between the membrane hole and other openings. 

\citeauthor{Tsai} (\citeyear{Tsai}) emphasizes that both the resonance frequency and damping coefficient of the membrane undergo changes before and after impedance measurement. Due to fluctuations in membrane tension over time, the measured resonance frequency could exhibit variations of up to 15\% between the initial and subsequent measurements. 
This variability has implications for accurately assessing the membrane's influence, revealing its inherent instability. Musicians playing the \textit{dizi} can distinctly perceive the instability in the membrane's condition during their performances. 
Therefore, the membrane parameters were empirically derived by comparing the measurements made with the membrane attached to the corresponding TMMI model results, with the best fit achieved using $f_m=$ \SI{4}{\kilo\hertz}, $m=$ \SI{8e-7}{\kilogram} and $R=$ \SI{3.5e-3}{\kilogram\per\second}. These values fall within the membrane parameter ranges reported in \citeauthor{Tsai} (\citeyear{Tsai}).

In subsequent analyses, the membrane hole without membrane (WOM) is treated as a standard closed hole (closed with Blu-Tack during measurements). When the membrane is attached (referred to as ``with membrane” or WM), the closed-hole shunt impedance is considered in series with the membrane impedance, assuming no mutual radiation between the membrane hole and other openings.





\subsection{Embouchure hole refinements}

\begin{figure}
    \centering
    \begin{minipage}{0.6\linewidth}
        \includegraphics[width=1\linewidth]{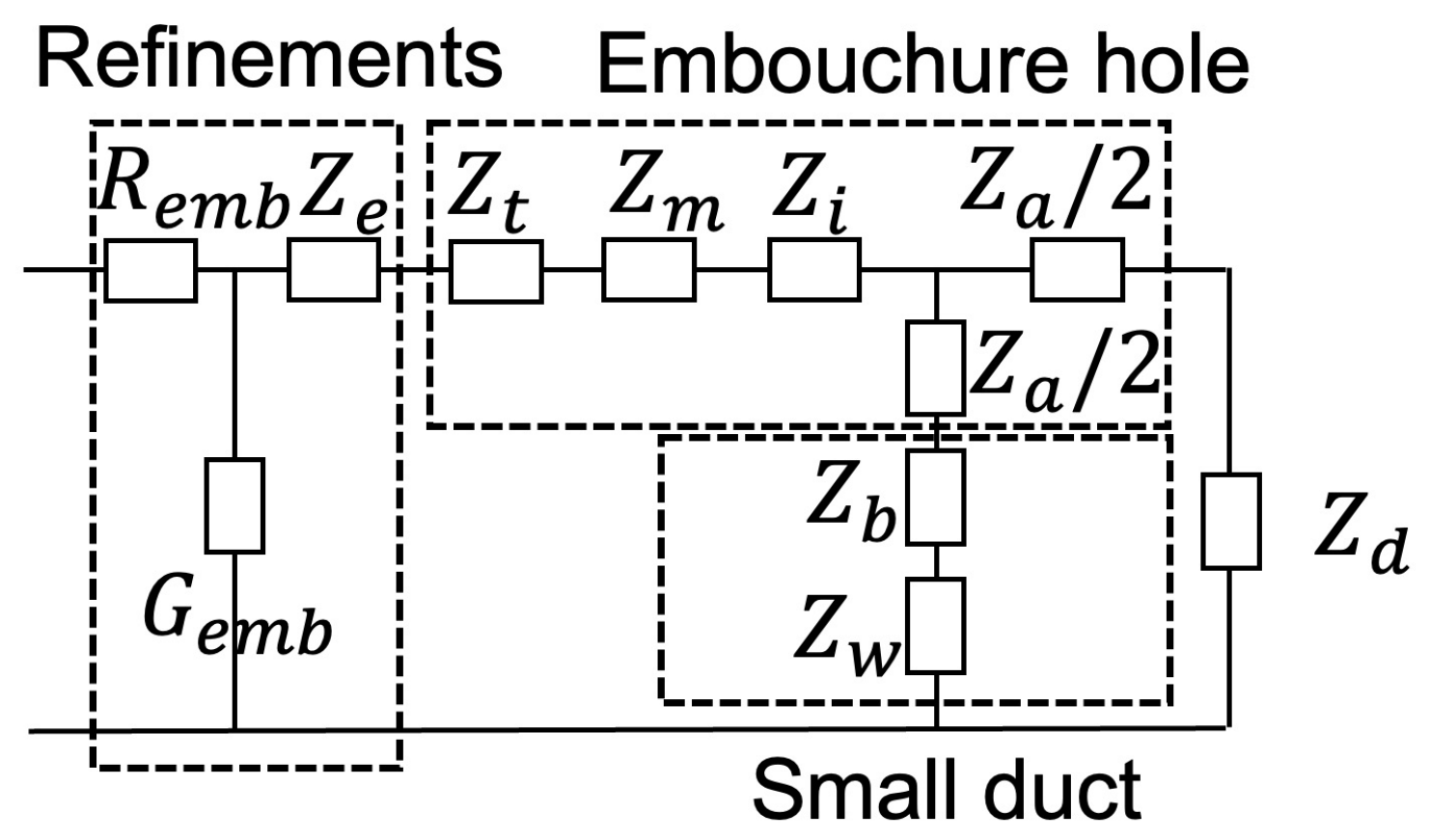}
    \end{minipage}%
    \begin{minipage}{0.32\linewidth}
        \caption{Equivalent circuits
representations for the refined model of the upstream branch coupled with the downstream.}
        \label{fig:emb}
    \end{minipage}
\end{figure}

The embouchure hole of the \textit{dizi} serves as the input, with the short duct between the cork and the embouchure hole and the downstream main duct connected in parallel.
In addition, the measurement system imposes a discontinuity at the flute headjoint and alters the measured input adminttance, as noted by \citeauthor{dickensthesis} (\citeyear{dickensthesis}). 
Therefore, a small length correction of $t_e = -$ \SI{1.7}{\milli\meter}  (associated with the impedance $Z_e$ shown in Fig.~\ref{fig:emb}) is applied to the embouchure hole to achieve closer fitting of admittance maxima. Apart from that, a series resistance $R_{emb} = (1 \times 10^{-5}$ Hz$^{-1})fZ_c $ and a shunt conductance $G_{emb} = (1 \times 10^{-4}$ Hz$^{-1}) f/Z_c$ are added at the input to achieve more accurate depth and height of admittance maxima and minima, respectively. 
Similar empirical refinements were employed by \citeauthor{dickensthesis} (\citeyear{dickensthesis}), assuming that the turbulence effects near discontinuities can be described as a dissipative process.
Figure \ref{fig:emb} provides an equivalent circuit representation for the refined model of the embouchure hole and the small duct coupled 
with the impedance of the downstream system. Together, we refer to the embouchure hole and the small duct as the upstream branch.
 $Z_b$ is the impedance of the small duct. $Z_w$ is the cork terminal impedance, which is approximated as being infinite, corresponding to a rigid wall. $Z_d$ is the impedance of the downstream main duct. 
 $Z_t$, $Z_m$, and $Z_i$ represent the impedances associated to 
the actual tonehole height $t$, matching volume
length correction $t_m$, and inner length correction $t_i$, respectively (see details in Appendix \ref{zazs}).

\subsection{Metrics}
In this section, we define several metrics that can be determined from admittances and used to help evaluate an instrument’s response or the performance of modeling by TMM and TMMI.

The frequencies that contribute to the sounding pitch of jet-blown instruments correspond to the maxima of the input admittance magnitude characteristics. Here, the frequency-dependent normalized admittance is defined in terms of admittance as $Y_{in}$. 
The frequencies of the admittance peaks influence the tuning, while their magnitudes are a factor that influences playability.
The harmonicity of the resonator is related to the frequency difference between the peak frequencies, given by 
\begin{equation}
    \Delta f = 1200 \times \log_2 \left (\frac{f_{n}}{nf_{1}}\right ),
    \label{eq: harmonicity}
\end{equation}
where $f_n$ is the evaluated peak frequency, $f_1$ is the first peak frequency, and
$\Delta f$ is measured in musical cents, with 1200 cents representing the value of one octave. The expected harmonicity for the ideal case is $\Delta f =0$.

The cutoff frequency of the oTHL can be estimated according to the reflection function \cite{petersen2020}. 
The reflection function can be seen as the transfer function between incident and reflected waves observed at the input of the resonator, which for admittance is defined as 
\begin{equation}\label{eq: R}
    {R} = \frac{{Y_{\textit{in}}}-1}{{Y_{\textit{in}}}+1}.
\end{equation}
Then the cutoff frequency is defined as the lowest frequency at which \cite{petersen2020}
\begin{equation}
    \frac{\left | R_{cutoff} \right |}{\left |{R}\right |_{\max}} =0.5.
    \label{eq: cutoff}
\end{equation}

Additionally, we utilize the normalized mean square error ($\operatorname{NMSE}$), expressed in $\si{\decibel}$, to assess the difference in admittance, along with the normalized cross-correlation ($\operatorname{NCC}$) to evaluate the similarity of admittance.
They are expressed by
\begin{equation}
    \operatorname{NMSE}(\hat{{Y}}_{{in}},{Y_{{in}}}) = 20\mathrm{log}_{10}\left ( \frac{{e}^H \cdot {e}}{Y_{in}^H \cdot {Y_{in}}} \right ),
\end{equation}
and
\begin{equation}
    \operatorname{NCC}(\hat{{Y}}_{{in}},{Y_{{in}}}) = \frac{\left |\hat{{Y}}_{{in}}^H \cdot {Y_{{in}}} \right |}{\|\hat{{Y}}_{{in}} \|_2 \cdot \| {Y_{{in}}}\|_2},
\end{equation}
where ${{{Y}}_{{in}}}$ are the measured data, $\hat{{Y}}_{{in}}$ are the simulated data, and ${e} = \hat{{Y}}_{{in}} - {Y_{{in}}}$ denote the error. Additionally, the metrics are computed with a column-vector representation of the data, and $\operatorname{NCC}$ reaches 1 when the two quantities match perfectly.
Note that both metrics operate on complex numbers, but their final results are real, and the superscript $H$ denotes the Hermitian transpose operator.

\section{Results\label{sec:Discussion}}


\subsection{\label{subsec:4:1} Comparison of Measurements and Models}


The main bore of the studied F-key \textit{bangdi} is not a perfect cylinder but has a slightly conical shape, with the downstream cross-section being slightly smaller than the upstream (see Table~\ref{table:borehole}). The geometry of the \textit{dizi} varies between instruments, as they are naturally crafted from bamboo.
We model the F-key \textit{bangdi}, both with and without the membrane, using the same fingerings as in the measurements.

\begin{figure}[b]
    \centering
    \includegraphics[width=1\linewidth]{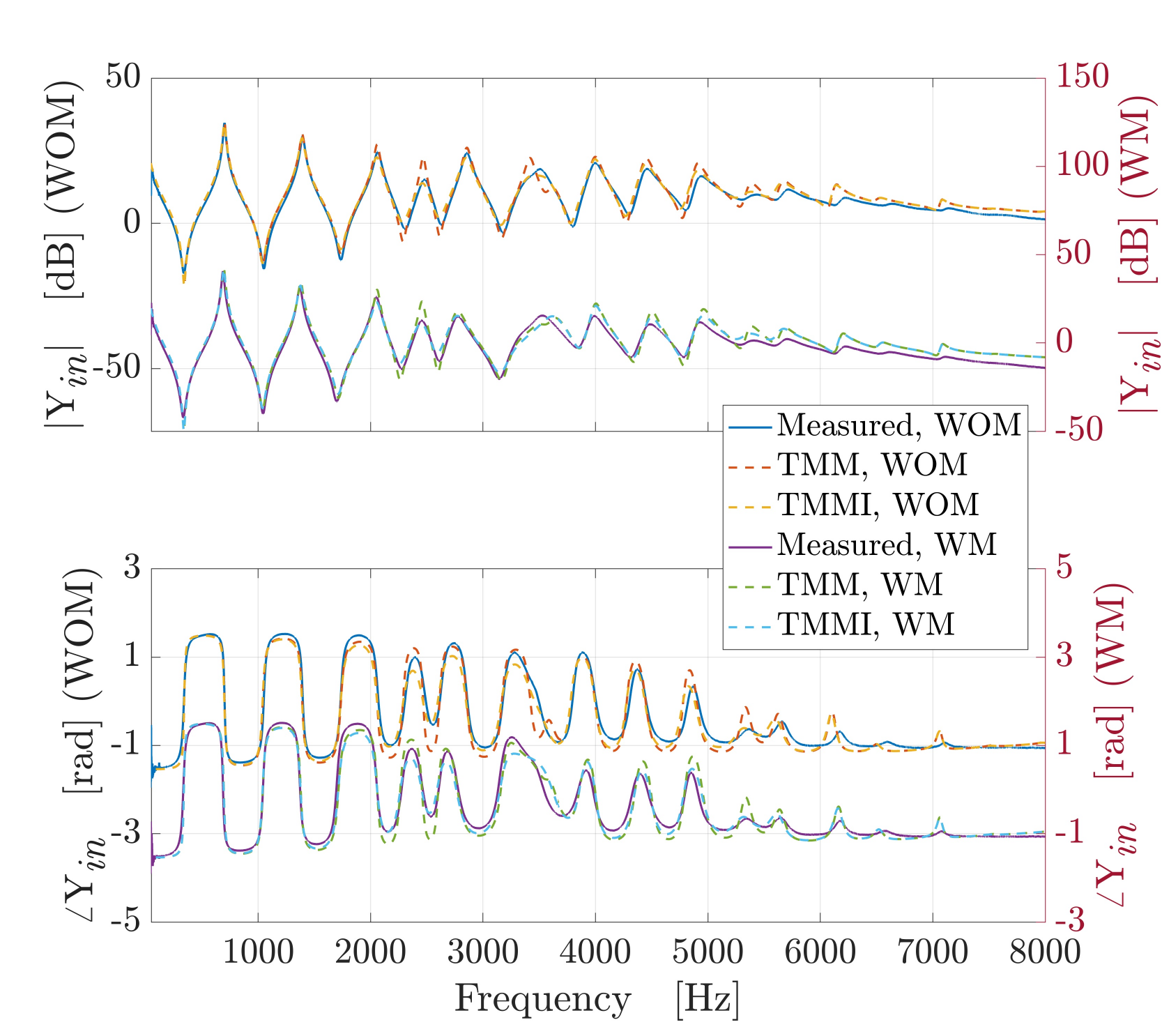}
    \caption{Normalized input admittance magnitude and phase of fingering XXXXOO, for WOM and WM.}
    \label{fig: xxxxoo_compare}
\end{figure}
\begin{figure}
    \centering
    \includegraphics[width=1\linewidth]{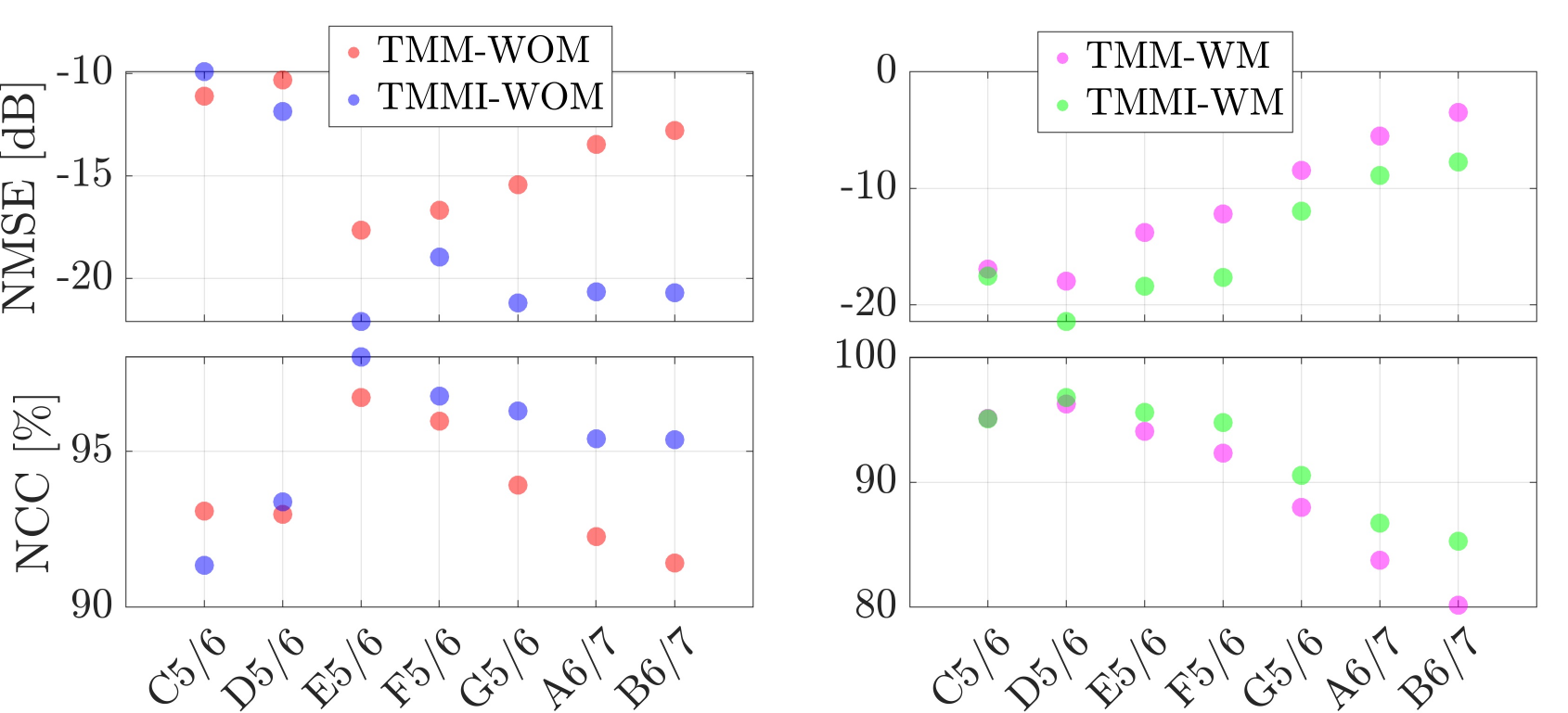}
    \caption{The NMSE (top) and NCC (bottom) of the input admittance for all fingerings, comparing TMM and TMMI with measurement results, for both WOM (left) and WM (right).}
    \label{fig:nmsencc}
\end{figure}

Figure~\ref{fig: xxxxoo_compare} shows an example result of normalized input admittance magnitude and phase for the fingering XXXXOO for both WM and WOM, obtained through measurement, TMM and TMMI.
From the observation, TMMI provides a more accurate representation of the peak shape around \SI{2.5}{\kilo\hertz} compared to TMM, indicating that the inclusion of mutual radiation impedance in TMMI is a valid enhancement.
A quantitative comparison of TMM and TMMI with the measurement results for all fingerings, both WM and WOM, based on the NMSE and NCC, is presented in Fig.~\ref{fig:nmsencc}. The frequency band used for analysis is $[100, 6000]$ \si{\hertz}. 
As expected, TMMI generally achieves higher accuracy in cases with more open holes, which is reflected in the results. For both WM and WOM, TMMI consistently shows lower NMSE and higher NCC values than TMM, except for the fingering XXXXXX. This exception may be attributed to an insufficient representation of the end-holes model, as the only open holes in this fingering are the end-holes. 
Moreover, the improvement is more pronounced in the absence of the membrane, indicating that the simplified membrane model may be insufficient.


\subsection{\label{subsec:membrane}Membrane Influence}
\begin{figure}
    \centering
\includegraphics[width=0.75\linewidth]{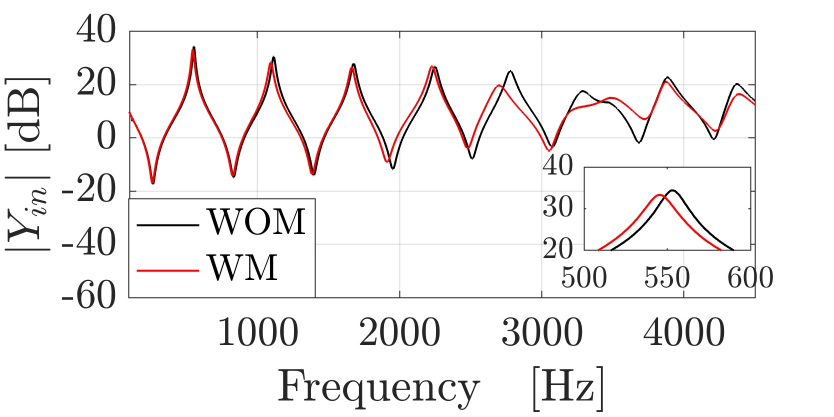}
  \caption{\label{memshift}Measured normalized admittance magnitude for XXXXXX.}
\end{figure}

As a particular component of the \textit{dizi}, the wrinkled membrane that covers the membrane hole is believed to contribute to the unique sound character of the \textit{dizi}.
To study the influence of the membrane from a linear perspective, the measured input admittance curves with and without the membrane are compared as shown in Fig.~\ref{memshift}, corresponding to the XXXXXX fingering. 

\begin{figure}
	\centering
	\includegraphics[width=1\linewidth]{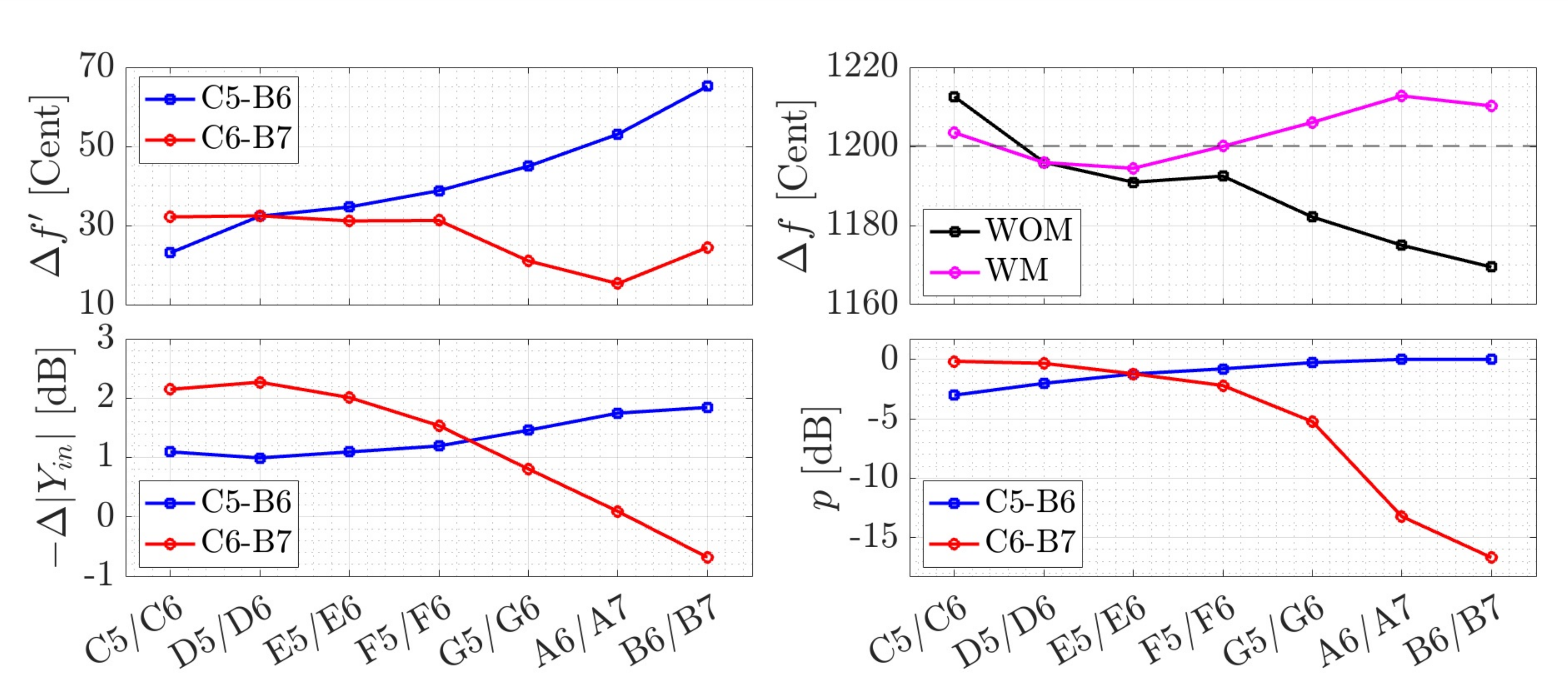}
	\caption{\label{membranevs}Top left: Frequency offset for the first two peaks due to the membrane.  
Top right: Frequency intervals of the peaks, for WM and WOM. 
Bottom left: Admittance magnitude shifts due to the membrane for the first two peaks.  
Bottom right: Normalized sound pressure at the membrane hole center, for the first two peaks. The first three graphs use measured data; the last graph is calculated using TMM. }
\end{figure}

 It can be seen that the maximum admittance peak values and their frequencies decrease for the first octave when the membrane is attached, which are referred to as  ``admittance reductions'' and ``resonance shifts'' \cite{Tsai}. The degree of the resonance shift and admittance reduction is related to the standing-wave pressure profile at the membrane hole location.
 Using a reference volume velocity value of one at the input (embouchure hole), the TMM is used to generate the pressure standing wave patterns to verify this relationship. Some results from this analysis are presented in Fig.~\ref{membranevs}.
By dividing the \textit{dizi} into several small segments (the length interval is less than \SI {1.2}{\centi \meter} for each segment), the relative pressure of each position can be calculated. 

Figure~\ref{membranevs} provides four different characteristics that can be used to assess the resonance shifts and admittance reduction effects due to the membrane. (A similar plot to Fig.~\ref{membranevs} can be found in Fig.~9 of \citeauthor{luanpoma} (\citeyear{luanpoma}). However, the legend labels for the second graph in Fig.~9 of \citeauthor{luanpoma} (\citeyear{luanpoma}) are reversed; please refer to the corrected legend provided in this paper.) 
The frequencies and magnitudes of the first two maxima in the admittance curve are extracted for all fingerings for both WM and WOM. 
From the overall comparison and analysis in Fig.~\ref{membranevs}, the curves for C5-B6 (the first peaks) show an upward trend in resonance shift (top left) and admittance reduction (bottom left), while the curves for C6-B7 (the second peaks) exhibit a downward trend in both. 
When examining the pressure levels at the membrane in the bottom right plot (with the pressure normalized by the maximum pressure along the bore for the corresponding frequency, on a logarithmic scale, so that the value of $0$ represents the maximum pressure), it is observed that the first peaks consistently show high pressure levels, with a slight increase in pressure as more finger-holes are opened. This is due to the fact that the first resonance peak generally features a pressure antinode at the center, positioning the membrane close to this region of maximum pressure. In contrast, the second peaks typically feature a pressure node at the center. As more finger-holes are opened, the membrane position shifts closer to this node, resulting in a noticeable reduction in pressure levels.
Based on the observations from the top left, bottom left, and bottom right plots, it reveals a positive correlation between the sound pressure level at the membrane hole and the extent of resonance shift and admittance reduction.
Additionally, from the top left plot, we observe that the membrane can shift the peak frequencies by 20-70 cents compared to when the membrane hole is sealed with Blu-Tack.
Note that \citeauthor{ng} (\citeyear{ng}) reported that a shift of less than 5 cents was found when they compared the membrane-attached hole to one sealed with stiff adhesive tape and this range falls within the just noticeable difference.
On the other hand, it can be found from the top right plot that the octave harmonicity (expected value: 1200 cents) shifts more for fingerings with more openings. 
Moreover, the peaks in the input admittance become more harmonically aligned when the membrane is attached, which suggests a potential improvement in tuning precision.
It is worth mentioning that those trends observed in Fig.~\ref{membranevs} could potentially impact playing behavior. However, further experiments with the \textit{dizi} being played would be necessary to gain a deeper understanding of the effects.

\subsection{\label{subsec:ubcutoff} Upstream branch and cutoff frequencies}

\begin{figure}
	\centering
	\includegraphics[width=1\linewidth]{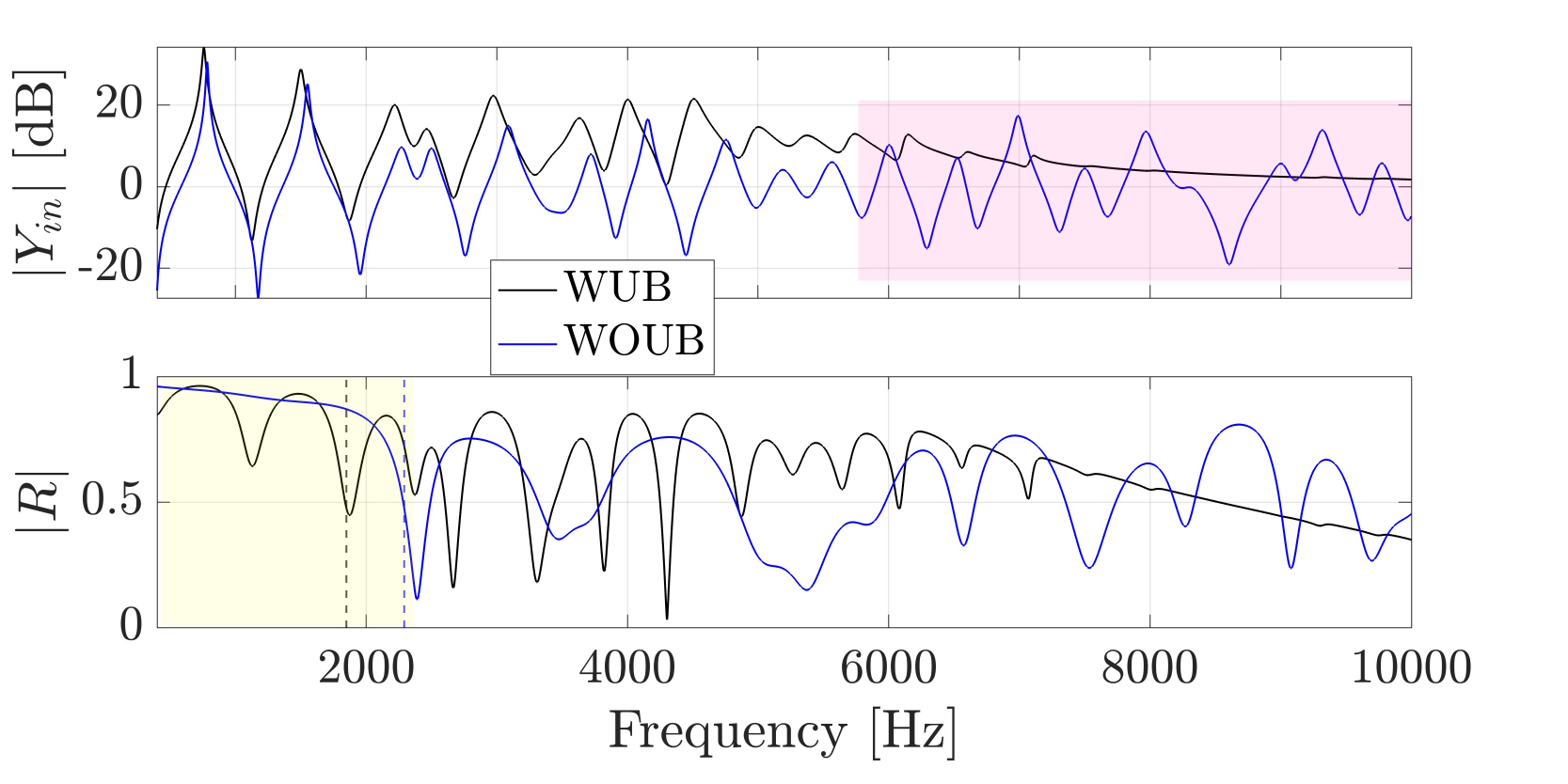}
	\caption{\label{ub}Normalized admittance magnitude and reflection coefficient magnitude for fingering XXXOOO (WOM).
 }
\end{figure}

The input admittance of the flute exhibits a sudden weakening around a specific frequency, attributed to the influence of the upstream branch as outlined in \citeauthor{Smith} (\citeyear{Smith}). This phenomenon, known as the Helmholtz shunt, similarly occurs in the \textit{dizi}.
In order to study the influence of the upstream branch,  Fig.~\ref{ub} compares the input admittance and reflection function curves modeled by TMMI, distinguishing between cases with the upstream branch (WUB) and without the upstream branch (WOUB) for the fingering XXXOOO, with no membrane attached. 
Looking at the region highlighted in pink, the upstream branch leads to a decrease in magnitude after \SI{5}{\kilo\hertz}, which is the Helmholtz shunt effect. 

\begin{figure}
	\figcolumn{
 \fig{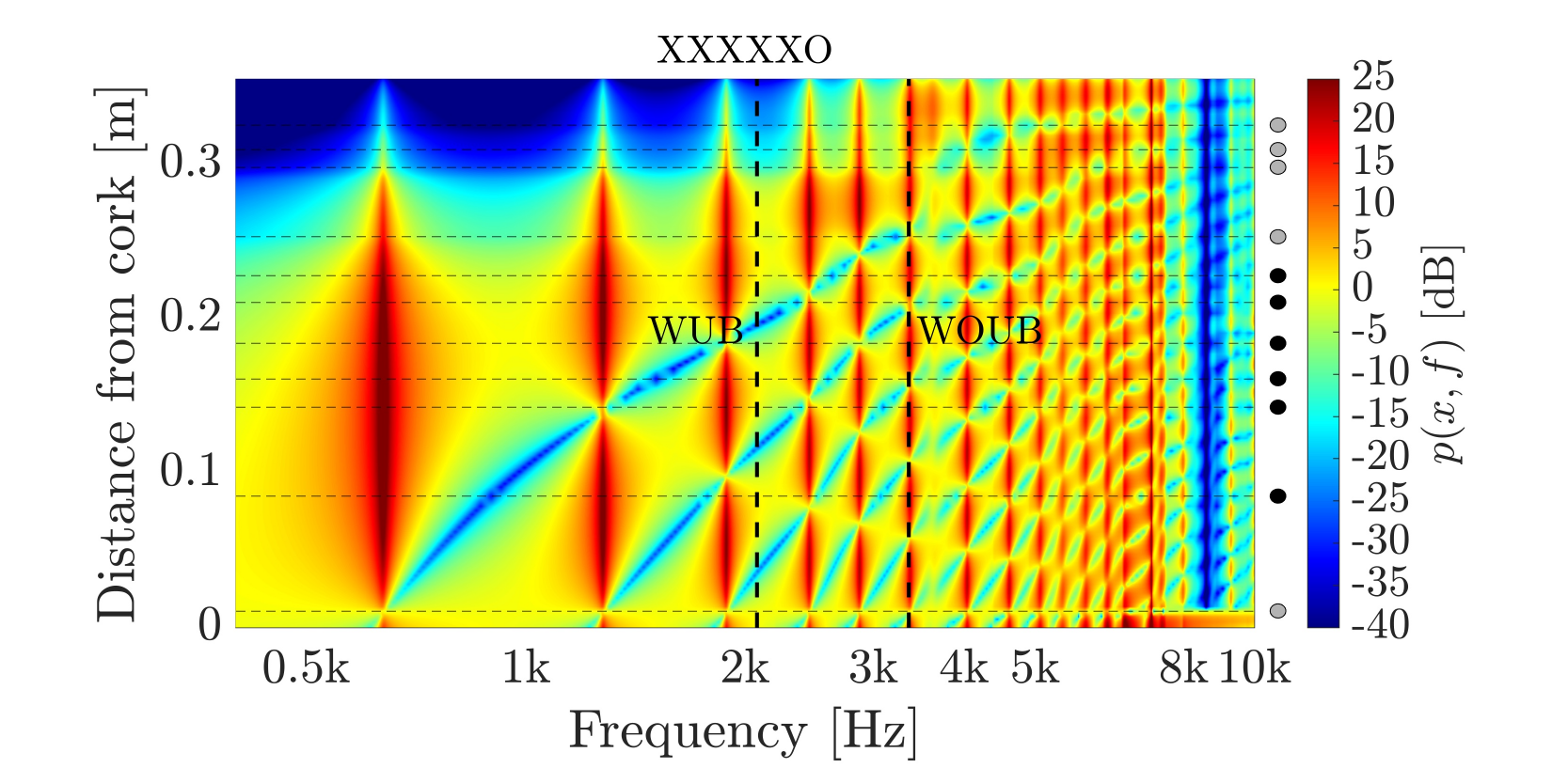}{1\linewidth}{}
\fig{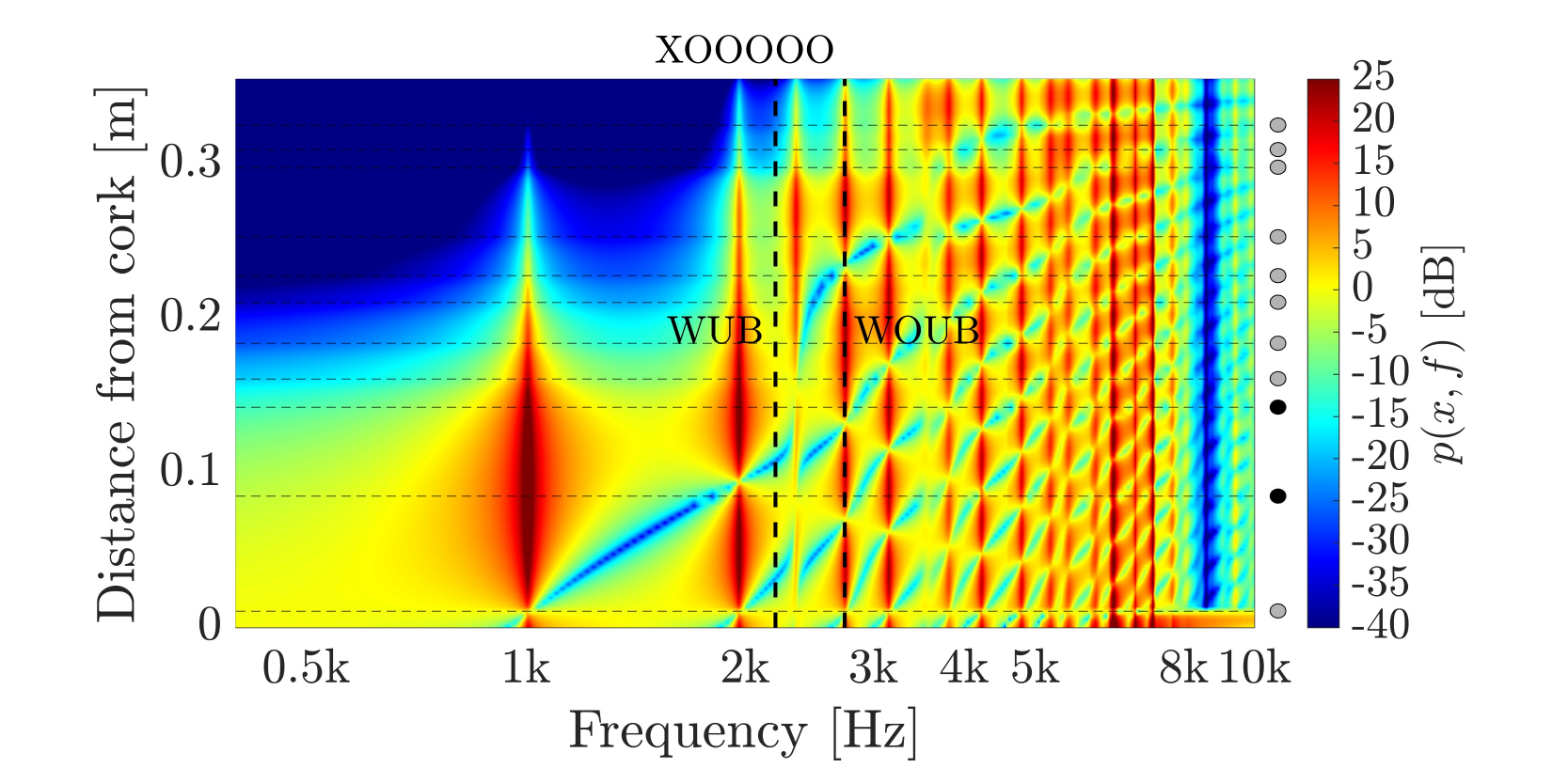}{1\linewidth}{}
    }
 \caption{Pressure frequency response to a unit impulse of volume velocity for WOM, for fingering XXXXXO (top) and XOOOOO (bottom). The frequency axis is in logarithmic scale. The cutoff frequencies extracted from WUB and WOUB using Eq.~\eqref{eq: cutoff} are indicated by dashed lines. Tonehole positions are marked by circles on the right, with grey for open holes.}
  \label{fig: pmap}
\end{figure}

The cutoff frequencies for the two cases, as determined from Eq.~\eqref{eq: cutoff}, are shown by dotted lines in the reflection function curves.
It can be seen that the upstream branch in flute instruments results in a more complex reflectance characteristic, which makes it harder to identify the cutoff frequency. 
To better understand the pressure distribution along the bore, we compute $p(x,f)$, the pressure response along the main bore (with $x=0$ at the embouchure hole), to a unit volume velocity excitation at the embouchure hole, $u(0,f) = 1$, using the TMM, based on the same algorithm as used for the bottom right plot in Fig.~\ref{membranevs}. This representation in the frequency domain can be interpreted as the impulse response in the time domain.
The spectrum maps of WOM, for fingerings XOOOOO and XXXXXO are shown as examples in Fig.~\ref{fig: pmap}. The cutoff frequencies that are found using Eq.~\eqref{eq: cutoff} are also marked in the figure. 
We know that the open tonehole lattice (oTHL) exhibits high-pass behavior, meaning that below the cutoff frequency, most sound cannot propagate to the downstream end \cite{benade1988clarinet, benade1988saxophone}. The pressure propagation behavior can be observed through the color changes in Fig.~\ref{fig: pmap}, which correspond to pressure levels in \si{\decibel}. 
Between the cutoff frequencies estimated from WUB and WOUB, all pressure components appear to decay significantly, by at least \SI{15}{\decibel}, before the end of the air column. On the other hand, all pressure components above the cutoff frequencies estimated from WOUB have significant levels extending all the way to the end of the instrument.
These analyses suggest that when evaluating the cutoff frequency of flute instruments, it is useful to remove the upstream branch so as to isolate the influence of the tonehole lattice.



    

\begin{figure}
    \centering
    \begin{minipage}{0.5\linewidth}
        \includegraphics[width=1\linewidth]{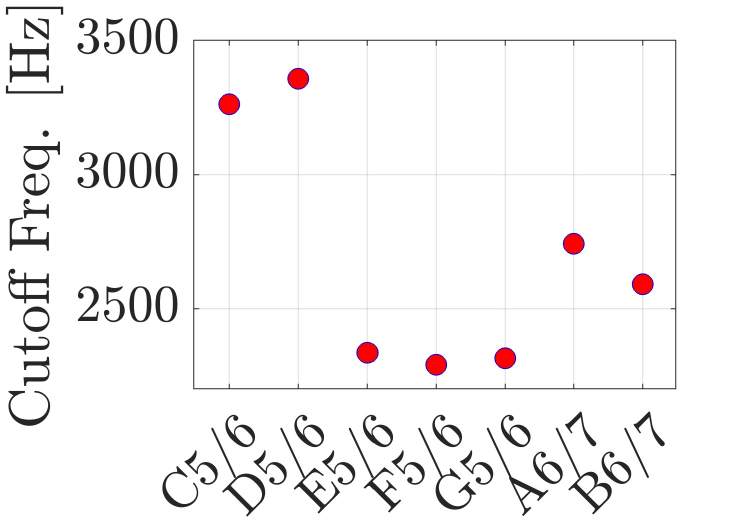}
    \end{minipage}%
    \begin{minipage}{0.3\linewidth}
        \caption{\label{fig: cutoff freq}The cutoff frequency from WOM \& WOUB, for all fingerings.}
    \end{minipage}
\end{figure}

We then extract the cutoff frequencies for all fingerings under both WOM and WOUB conditions, as shown in Fig.~\ref{fig: cutoff freq}. Notably, the cutoff frequencies for the first two fingerings are significantly higher than for the others. We suspect that this distinction in cutoff frequency behaviour results from the interaction between the four always-open end-holes, or end-hole lattice (EHL), and the finger-hole lattice (FHL). Examining Fig.~\ref{fig: pmap}, we observe that for the XOOOOO fingering (many open finger-holes), the pressure components below the WOUB cutoff frequency are primarily influenced by the FHL, whereas for the XXXXXO fingering (most finger-holes closed), the pressure components below the WOUB cutoff frequency propagate up to and into the EHL. Given the different hole spacings in the two lattices, as well as the distance between the EHL and FHL, it is reasonable to expect the observed difference in cutoff frequencies shown in Fig.~\ref{fig: cutoff freq}. We also expect that the difference in cutoff frequencies will influence the \textit{dizi}’s radiation and tonal characteristics, or timbre, though the extent of this impact and its perceptual significance remain topics for future investigations.

\subsection{\label{subsec:allyin} Input admittance analysis}

\begin{figure*}
    \centering
    \includegraphics[width=\linewidth]{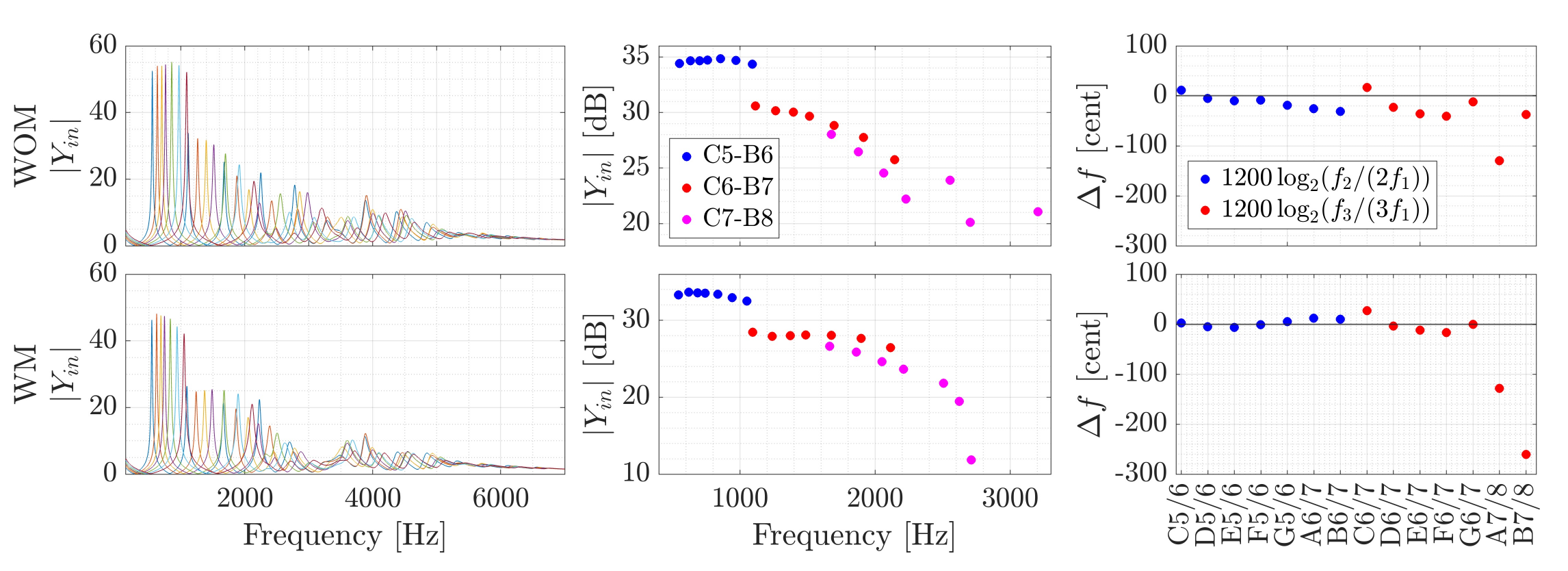}
        \caption{Left: Normalized input admittance (linear scale) for all fingerings. Middle: Frequency and magnitude (in \si{\decibel}) of the first three peaks of the normalized input admittance. Right: Harmonicity between the first three peaks. Top: WOM. Bottom: WM. All results are based on measurement data.}
        \label{fig:zin_all_desc}
\end{figure*}

In order to analyze the overall characteristics of the \textit{dizi}, the measured input admittance magnitudes, as well as the extracted peak frequencies, magnitudes, and harmonicities for all fingerings, both WM and WOM, are shown in Fig.~\ref{fig:zin_all_desc}.
Looking at the left panels of Fig.~\ref{fig:zin_all_desc}, the F-key \textit{bangdi} can be divided into three frequency bands according to the characteristics of the input admittance curve: less than  \SI{2.3}{\kilo\hertz}, 2.3-\SI{5}{\kilo\hertz}, and greater than 5 \si{\kilo\hertz}.
In the first band below \SI{2.3}{\kilo\hertz}, the envelope of the input admittance curve drops relatively slowly.
The frequency of the second band is 2.3 to \SI{5}{\kilo\hertz}, and the peak values of the admittance are relatively reduced, which can be explained as the effect of the cutoff frequency of the oTHL, and agrees well with the cutoff frequency results in  Fig.~\ref{fig: cutoff freq}. 
The irregularity around  \SI{3.5}{\kilo\hertz} is due to the resonant frequency of the membrane, as explained in Sec.~\ref{subsec:membrane}. 
In the third band above \SI{5}{\kilo\hertz}, the input admittance curve suddenly becomes flat for each fingering. This behavior is characteristic of flute instruments and is related to the Helmholtz shunt effect, as explained in  Sec.~\ref{subsec:ubcutoff}.
The admittance peak frequencies are depicted in the middle panels, while the harmonicity between the  first and second, and the second and third peaks, computed according to Eq.~\eqref{eq: harmonicity}, is presented in the right panels. 
The results indicate that the membrane affects the height of the admittance peak that sustains the oscillation for the fingering OOOOOO, as its frequency is near the resonance frequency of the wrinkled membrane. 
Additionally, the membrane appears to affect the harmonicity of A7/A8 and B7/B8, which also lie near the membrane's resonance frequency. Overall, however, the presence of the membrane generally improves harmonicity (see the right panels).

\section{\label{sec:conclusion}Conclusion}
In this study, we investigate the linear acoustical behavior of the \textit{dizi} resonator. An acoustical model of each component, including the drilled toneholes, back end-holes (two radially positioned holes), membrane hole, and the coupling with the upstream branch of the \textit{dizi}, is presented. The input admittance of the \textit{dizi} is measured and modeled using TMM and TMMI, cascading the acoustical models of each component.
The comparison between TMMI, TMM, and the measured data indicates reasonable modeling accuracy. TMMI outperforms TMM in both NMSE and NCC, due to its consideration of external radiation interactions. Since the geometry of each \textit{dizi} varies, this acoustical modeling can serve as a useful tool for \textit{dizi} makers, allowing them to assess the tuning of each instrument by inputting its specific geometric parameters.

Acoustical analyses of the \textit{dizi} resonator are then conducted, considering each individual component and the entire instrument as a whole.
The F-key \textit{bangdi} can be divided into three frequency bands according to the characteristics of the input admittance
curve: less than 2.3 \SI{}{\kilo\hertz}, 2.3-5 \SI{}{\kilo\hertz}, and greater than 5 \SI{}{\kilo\hertz}. 
Attaching the membrane to the \textit{dizi} shifts the admittance curve peaks to lower frequencies and reduces their magnitude. The membrane’s influence on tuning and harmonicity varies depending on the fingering. In general, its presence enhances harmonicity; however, when the frequency approaches the resonance of the wrinkled membrane, the resulting interaction can lead to deviations in both harmonicity and tuning accuracy. 
It is possible that instrument makers are aware of this effect and take it into consideration when designing and positioning the tone holes, potentially leveraging the membrane's impact to enhance tuning accuracy and harmonicity.

When evaluating the cutoff frequency of flute instruments, the upstream branch can be removed in order to more clearly understand the influence of the tonehole lattice.
The cutoff frequencies exhibit two distinct groups across different fingerings, with the first two fingerings being significantly higher than the others. This is likely due to the presence of two oTHLs: the FHL and EHL. 
Additionally, we suggest that greater attention may need to be given to
the acoustic function of the end-holes, as they may influence the \textit{dizi}'s radiation pattern and timbre.


Future work could focus on developing more accurate numerical models for cylindrical mutual radiation and back-end holes. Additionally, a geometry optimization algorithm based on input admittance could assist in instrument design and manufacturing. The current acoustical model may also be applied to physics-informed sound synthesis of the \textit{dizi}. Furthermore, a quantitative analysis of tone color through radiated sound across different fingerings could enhance our understanding of the FHL and EHL.
Another interesting direction is to study the nonlinear dynamics of the membrane, which is a distinctive feature of the \textit{dizi} and plays a crucial role in shaping its characteristic timbre.

\section*{Acknowledgments}
This work was partly supported by National Social Science Fund of China 22VJXG012 and 22\&ZD037.

\section*{Author declarations}
\subsection*{Conflict of Interest}
The authors have no conflicts to disclose.

\subsection*{Data availability}
The data that support the findings of this study are available within the article.

\appendix
\section{}
\subsection{Dimension of the \textit{bangdi} in F}\label{app:dim}

Table \ref{table:borehole} provides the dimensions of the F-key \textit{bangdi} used in this study, for the bore and toneholes, respectively. 
 The center of the embouchure hole is regarded as the origin of the axial position, with the direction from the upstream to the downstream as the positive direction. 
The sequence of the toneholes is: embouchure hole, membrane hole, 6 finger-holes, back end-holes, 2 front end-holes. 
It is worth noting that the dimensions for the back end-holes are for a single hole, as the shape of the two holes is identical with the same axial position. The bore radii are determined from their diameters. Since the tonehole surfaces are not perfect circles, we measure the lengths of the semi-major axis (multiplied by 2) and the semi-minor axis (multiplied by 2) of the ellipse and calculate the equivalent radius of a circle by ensuring that the areas are equal.

\begin{table*}[ht]
{\fontsize{8}{10}\selectfont
\caption{Bore and tonehole dimensions: $x$ is the axial coordinate, $a$ the bore radii, and $b$ the tonehole radii.}
\begin{tabular}{ccrrrrrrrrrrrrrrrr}
\hline
\hline
 Bore & {\shortstack{$x$}} & $-10.6$ &0 & 17.4& 42.5 &57.3  & 82.5 & 98.2  & 120.3 & 163.9& 187.4 & 232.0 & 257.8 & 330.4 &  334.3 & 334.3& 344.8 \\
 (\si{\milli\meter}) & {\shortstack{$a$}}
&7.2 &7.2  &  7.2  &  7.3 &   7.3  &  7.1 &  7.1 & 7.1 &  7.0 &  7.0 &  6.8 &  6.6  &  5.8 & 6.2 &   5.8& 5.9\\
\hline
 Tonehole & {\shortstack{$x$}} &0 &74.4 & 132.1 & 150.5  &173.5 & 200.1 &217.3 & 242.5 &287.6 & 299.0 &314.9\\
  (\si{\milli\meter})& {\shortstack{$b$}}
&4.7 &3.9  &  4.4 &  4.4  & 4.4 & 4.3 &   4.5 & 4.3  &  4.4 & 4.3 &  4.4   \\
\hline
\hline
\multicolumn{18}{l}{$^*$The height of all toneholes, $t$, is \SI{4}{\milli\meter}.}\\
\end{tabular}
\label{table:borehole}
}
\end{table*}

\section{}
\subsection{Propagation constant in a cylindrical tube}\label{const}

\subsection{Various impedances in TMM}\label{zazs}

The series impedance $Z_a$ of the open and closed tonehole can be regared as a small negative acoustic mass
\begin{equation}
	Z_a^{(c/o)}=j \tan(kt_a^{(c/o)})Z_c\approx jkt_a^{(c/o)}Z_c,
\end{equation}
where the superscripts $c$ and $o$ stand for closed and open, respectively, $k=2\pi f/c$ is the lossless wave number, $f$ is the frequency, $Z_c = \rho c /\pi a^2$ is the characteristic impedance of the bore radius $a$,  $\rho$ and $c$ represent the density and the velocity of sound in air.

The calculation of the shunt impedance $Z_s$ can be relatively complicated.
For the closed hole, it is mainly represented by the acoustic compliance \cite{nederveen1969}, given by \citeauthor{nederveen} (\citeyear{nederveen}),
\begin{equation}
	Z_s^{(c)}=j\{kt_i-\cot[k(t+t_m)]\}Z_c,
\end{equation}
where $t_i$ and $t_m$ are the inner length and matching volume length correction, respectively.

Compared with the closed hole, $Z_s^{(o)}$ includes the radiation impedance, which can be expressed by the radiation length correction $t_r$ \cite{dalmont},
\begin{equation}
	Z_s^{(o)}=j\{kt_i+\tan[k(t+t_m+t_r)]\}Z_c.
\end{equation}

The expression of $t_a, t_i$ is given from \cite{Lefebvre2012},
\begin{equation}
	t_a^{(c)}=[-0.12-0.17\tanh(2.4t/b)]b\delta^2,
\end{equation}

\begin{equation}
	t_a^{(o)}=[-0.35+0.06\tanh(2.7t/b)]b\delta^2,
\end{equation}
with $\delta=b/a$, and $b$ the tonehole radius.
The variable $t_i$ can be expressed with a multiplicative factor $G(\delta,ka)=[1+H(\delta)I(ka)]$ to account for frequency dependence,
\begin{equation}	
\begin{split}
t_i^{(c/o)}=(0.822-0.095\delta-1.566\delta^2+ 2.138\delta^3  -1.640\delta^4+0.502\delta^5)b G(\delta,ka),
\end{split}
\end{equation}
where
\begin{equation}
	H(\delta)=1-4.56\delta+6.55\delta ^2,
\end{equation}
and
\begin{equation}
I(ka)=0.17ka+0.92(ka)^2+0.16(ka)^3-0.29(ka)^4.
\end{equation}

The matching volume length correction of the unflanged hole is shown here \cite{nederveen}
\begin{equation}\label{tm}
	t_m^{(o/c)}=\frac{b\delta}{1+0.207\delta^3},
\end{equation}
while the $t_m$ used in the modeling of the \textit{dizi} in this manuscript follows Eq. (\ref{eq:match}).

The radiation length correction of cylindrical flanges is given in \citeauthor{dalmont2001radiation} (\citeyear{dalmont2001radiation}) and \citeauthor{Lefebvre2012} (\citeyear{Lefebvre2012}), which could be used for toneholes drilled through a thick wall,
\begin{equation}
    t_r = 0.8216 b -0.47 b [b/(a+t)]^{0.8}.
\end{equation}

Since the wall thickness of the \textit{dizi} is non-negligible, the end correction $Z_r^{L}$ for a tube with an infinite flange is derived in 
\citeauthor{norris} (\citeyear{norris}) and \citeauthor{dalmont2001radiation} (\citeyear{dalmont2001radiation}).
\begin{equation}
\begin{split}
    &Z_r^{L} = [(1+R_c)/(1-R_c)] Z_c, \\
    &R_c = - |R_{\infty}| e^{-2jk\delta _{\infty}}, \\ 
    & \delta _{\infty} = 0.8216a \left[1+ \frac{(0.77ka)^2}{1+0.77ka}\right ] ^{-1},\\
    & |R_{\infty}| = \frac{1+0.323ka-0.077(ka)^2}{1+0.323ka+(1-0.077)(ka)^2}.
    \end{split}
    \label{eq:endrad}
\end{equation}

\subsection{Matrix $\mathbf{B}$, $\mathbf{Z}$ and $\mathbf{Y}$ in TMMI}\label{app:tmmi}

$\mathbf{B}$ stands for the acoustic mass of the open holes, which is a diagonal matrix, corresponding to the impedance with the length $t$ and length corrections $t_i$ and $t_m$. Its $i$th diagonal element is 
\begin{equation}
    B_{ii} =j\{kt_{i,ii}+\tan[k(t_{ii}+t_{m,ii})]\}Z_{c,ii}.
\end{equation}

The radiation matrix $\mathbf{Z}$ includes both the self-radiation and mutual radiation impedance, with the diagonal elements representing self-radiation and the off-diagonal elements representing mutual radiation.

The self-radiation impedance of the $i$th opening for a tonehole is
\begin{equation}
    Z_{ii} =j\tan(k t_{r,ii})Z_{c,ii}.
\end{equation}
The self-radiation impedance of the end ($i=N$) follows Eq. (\ref{eq:endrad}), $Z_{ii} = Z_r^{L}$.

By assuming the open ends radiate as monopoles, the mutual radiation impedance $Z_{ij}$ (when $i \neq j$) is \cite{pritchard}
\begin{equation}
	Z_{ij}=jk\rho c\frac{e^{-jkd_{ij}}}{ \epsilon \pi d_{ij}},
\end{equation} 
where $d_{ij}$ is the distance between open ends $i$ and $j$, $ \epsilon$ is a factor corresponding to the radiation space, $\epsilon=2$ for a half space, and $\epsilon=4$ for a complete space. As mentioned in  \citeauthor{tmmi} (\citeyear{tmmi}), empirically it is difficult to determine the best approximation for the radiation impedance. They suggest to use $\epsilon=2$ when the effect of interaction is especially important. In this manuscript, we use $\epsilon=2$ between all axially distributed holes, and $\epsilon=4$ between axially distributed holes and the end.

However, given the unique structure of the \textit{dizi} with two radially distributed back end-holes, we need to define the mutual radiation impedance differently. We treat the back end-holes as a single component, as discussed in Sec. \ref{subsec:3:3}. In this scenario, the mutual radiation impedance between these holes and other openings is increased due to the double radiation area; hence, we assume $\epsilon=1$. 
Additionally, the mutual radiation between them is not taken into account because their mutual radiation effects have already been effectively considered through self-radiation.
This implies that, from a physical standpoint, the shunt impedance includes contributions from both self-radiation and mutual radiation.

The admittance matrix $\mathbf{Y}$ is related to the TM between two openings, corresponding to all the cascaded components between the two holes' shunt impedance $Z_s$ or the end radiation impedance $Z_r^L$.

\begin{equation}
	\begin{bmatrix}
	p_i \\ u_i
\end{bmatrix} = \begin{bmatrix}
	A_i &B_i\\ C_i & D_i
\end{bmatrix}\begin{bmatrix}
	p_{i+1}\\u_{i+1}	
\end{bmatrix},
\label{eq:y1}
\end{equation}
which can be written in the form of an admittance matrix 
\begin{equation}
    \begin{bmatrix}
        u_i \\ u_{i+1} 
    \end{bmatrix} = 
    \begin{bmatrix}
        Y_i & Y_{\mu,i} \\ Y_{\mu,i} & Y_i'
    \end{bmatrix}
    \begin{bmatrix}
        p_i \\ p_{i+1}
    \end{bmatrix},
\end{equation}
where the relationship between the parameters is given by: $Y_i = D_i/B_i$, $Y_i' = A_i/B_i$ and $Y_{\mu,i}=-1/B_i$, which assumes that $A_i D_i -B_i C_i=1$, the condition for reciprocity.

\bibliography{sampbib}

\end{document}